\documentclass[12pt]{article}
\usepackage{graphicx}

\begin{document}
\newcommand{\bfm}[1]{\mbox{\boldmath{$#1$}}}
\newcommand{\beq}{\begin{eqnarray}}
\newcommand{\eeq}{\end{eqnarray}}

\date{Submitted to {\it Icarus} May 2, 2008}

\title{Effect of Density Inhomogeneity on YORP:  The case of Itokawa}

\author{D.J. Scheeres\\
Colorado Center for Astrodynamics Research\\
The University of Colorado at Boulder\\
\\
R.W.\ Gaskell\\
Planetary Science Institute}

\bigskip
\maketitle

%\end{opening}
%\begin{document}

\newpage
\noindent {\bf Please, send editorial
communications and proofs to:} \\ Dan Scheeres\\ 429 UCB \\
Boulder, CO 80309-0429\\ E--mail: scheeres@colorado.edu\\

\bigskip \bigskip \bigskip \bigskip \bigskip \bigskip
\noindent
Manuscript pages: 15
\medskip

\noindent
Figures: 4
\medskip

\noindent
Running title: Effect of Density Inhomogeneity on YORP
\medskip

\newpage

\baselineskip=2\baselineskip
%\baselineskip=\baselineskip

\begin{abstract}
The effect of density inhomogeneity on the YORP effect for a given shape model is investigated.  A density inhomogeneity will cause an offset between the center of figure and the center of mass and a re-orientation of the principal axes away from those associated with the shape alone.  Both of these effects can alter the predicted YORP rate of change in angular velocity and obliquity.  We apply these corrections to the Itokawa shape model and find that its YORP angular velocity rate is sensitive to offsets between its center of mass and center of figure, with a shift on the order of 10 meters being able to change the sign of the YORP effect for that asteroid.  Given the non-detection of YORP for Itokawa as of 2008, this can shed light on the density distribution within that body.  The theory supports a shift of the asteroid center of mass towards Itokawa's neck region, where there is an accumulation of finer gravels.  Detection of the YORP effect for Itokawa should provide some strong constraints on its density distribution.  This theory could also be applied to asteroids visited by future spacecraft to constrain density inhomogeneities.
\end{abstract}

\section{Introduction}

The YORP effect for asteroids, initially proposed by Rubincam \cite{YORP_rubincam}, has now been verified \cite{YORP_lowry, YORP_taylor, YORP_Apollo} and is being invoked to explain a variety of evolutionary dynamics and states for the smaller members of the asteroid population.  Despite this success, the YORP effect for the one small NEA whose shape and mass is best characterized, the asteroid Itokawa \cite{gaskell_1, gaskell_2, science_abe}, has not been detected yet, despite predictions that YORP should have been evident by now \cite{Icarus_itokawa}.\footnote{It should be noted that a detection of YORP for Itokawa was reported in \cite{kitazato}, but was subsequently retracted by that author at the 2007 DPS meeting.}  The non-detection of the YORP effect for Itokawa raises a number of interesting modeling questions about the details of the YORP effect.  

Rubincam noted that density inhomogeneity could affect computed YORP torques \cite{YORP_rubincam}, although this has not been explicitly investigated in the literature. 
In the following we will show that density inhomogeneity within an asteroid has a direct influence on the predicted YORP spin rate evolution, and present the theory to compute this effect.  
We present computations of the change in the YORP effect for high-precision shapes of Itokawa.  We find that center of mass shifts on the order of a few meters can significantly change the YORP spin-rate deceleration, and that shifts on the order of 10-15 meters could change its sign and make Itokawa's spin rate accelerate.  The current non-detection of YORP for Itokawa is consistent with a density concentration in the neck region of Itokawa.  Once the YORP effect for Itokawa is detected, the theory presented here can help us constrain its density inhomogeneity.  Similar YORP measurements for asteroids to be visited in the future could also provide insight into their density distributions.

\section{Theory}

At any given instant the force and moment acting on an asteroid due to incident, reflected and reemitted solar radiation from its surface can be represented as a function of the sun's location in an asteroid-fixed reference frame.  In \cite{scheeres_YORP} the representation of these forces and moments using a Fourier series expansion is described in detail, and algorithms for the computation of the coefficients, starting from a polyhedral shape model, are given.  The direction of the sun in the asteroid body-fixed frame can be described by a unit vector, $\hat{\bfm{u}}$.  For a given solar location we can compute the total force and moment acting on the asteroid due to the sunlight:
\beq
	\bfm{F}(d, \hat{\bfm{u}}) & = & \frac{{\cal G}_1}{d^2} \sum_{i=1}^N \bfm{f}_i(\hat{\bfm{u}}) \\
	\bfm{M}(d, \hat{\bfm{u}}) & = & \frac{{\cal G}_1}{d^2} \sum_{i=1}^N \bfm{r}_i\times\bfm{f}_i(\hat{\bfm{u}}) 
\eeq
where $\bfm{F}$ and $\bfm{M}$ are the total force and moment acting on the asteroid at a given instant, ${\cal G}_1\sim 1\times10^{14}$ kg km/s$^2$ is the solar constant, $d$ is the distance from the asteroid to the sun, $\bfm{f}_i$ is the force acting on each surface facet, both at the instant of insolation and after a characteristic time-lag for reemission, $\bfm{r}_i$ is the position vector from the presumed center of mass of the body to the surface facet, and $N$ is the number of facets for the shape model.  When expressed as a Fourier series (or as a Spherical Harmonic series in \cite{CMDA_YORP}) these forces can be represented as:
\beq
	\bfm{F}(d, \hat{\bfm{u}}) & = & \frac{{\cal G}_1}{d^2} \sum_{l=0}^{\infty} \left[ \bfm{A}_l(\delta) \cos(\lambda - \theta) + \bfm{B}_l(\delta)\sin(\lambda-\theta)\right] \\
	\bfm{M}(d, \hat{\bfm{u}}) & = & \frac{{\cal G}_1}{d^2} \sum_{l=0}^{\infty} \left[ \bfm{C}_l(\delta) \cos(\lambda - \theta) + \bfm{D}_l(\delta)\sin(\lambda-\theta)\right]
\eeq
where $\delta$ is the solar latitude, computed from $\sin\delta  = \sin i \sin(\varpi + \nu)$, $\lambda$ is the solar longitude, $\theta$ is the asteroid rotation angle about its rotation pole, $i$ and $\varpi$ are the asteroid orbit elements of inclination and argument of periapsis in the asteroid-fixed frame with the $z$-axis taken as the rotation pole, and $\nu$ is the true anomaly.  The Fourier coefficients are shown as being functions of the latitude, however they can be further expanded in constant coefficients using Associated Legendre Functions \cite{CMDA_YORP}.  The Fourier coefficients for the moment, $\bfm{C}_l$ and $\bfm{D}_l$, are directly computed from the coefficients for the force, where we note that the total coefficient equals the sum of the coefficients for each facet.  Thus we have the relationships:
\beq
	\bfm{C}_l & = & \sum_{i=1}^N \bfm{r}_i \times \bfm{A}_l^i \\
	\bfm{D}_l & = & \sum_{i=1}^N \bfm{r}_i \times \bfm{B}_l^i 
\eeq
where $\bfm{A}_l^i$ and $\bfm{B}_l^i$ denote the Fourier coefficients at order $l$ for the $i$th face.  In these expressions we suppress the time-lag, although it can be included into this general form.

As described in \cite{scheeres_YORP, CMDA_YORP}, the rotational dynamics of an asteroid subject to these torques can be averaged over one rotation about its axis and about one revolution around the sun.  Doing so analytically removes almost all of the moment coefficients from consideration, and isolates the secular contributions of YORP to the asteroid spin rate, $\omega$, obliquity, $\epsilon$, and right ascension, $\alpha$:
\beq
	\dot{\omega} & = & \frac{G_1}{I_z a^2 \sqrt{1-e^2}}  \bar{C}_{0,z}(\epsilon) \label{eq:ommsecE} \\
	\dot{\epsilon}  & = & \frac{G_1}{\omega I_za^2 \sqrt{1-e^2}} \label{eq:ipmsecE} \\
	& & 
	\left[ \left( \bar{C}_{1,x}(\epsilon) + \bar{D}_{1,y}(\epsilon) \right) \cos(\phi_{lag}) + \left( \bar{D}_{1,x}(\epsilon) - \bar{C}_{1,y}(\epsilon) \right) \sin(\phi_{lag}) \right]  \nonumber \\
	\dot{\alpha}  & = & - \frac{\cot(\epsilon) G_1}{\omega I_za^2 \sqrt{1-e^2}} \label{eq:opmsecE} \\
	& & 
	\left[ - \left( \bar{D}_{1,x}(\epsilon) - \bar{C}_{1,y}(\epsilon) \right) \cos(\phi_{lag}) + \left( \bar{C}_{1,x}(\epsilon) + \bar{D}_{1,y}(\epsilon) \right) \sin(\phi_{lag}) \right] \nonumber
\eeq
where $\phi_{lag}$ is the thermal lag angle, and $a$ and $e$ are the asteroid orbit's semi-major axis and eccentricity.  This form of the secular equations implicitly assumes that only the reemission terms are present.  

\section{Corrections due to density inhomogeneities}

A common assumption in all previous analyses of YORP effects for given shape models, including the theory given above, has been that the body has a uniform density distribution, and thus that the $\bfm{r}_i$ are measured from the center of figure for a given shape.  In the following we continue to assume that the moments and forces are computed from the center of figure of the body, but now we will also allow for the center of mass and the principal axis orientations of the body to be different from the homogeneous density case.  The theory developed in \cite{scheeres_YORP, CMDA_YORP} has the common assumption that the dynamics are evaluated relative to the center of mass of the system and in a principal axis frame.  We will continue with these assumptions and, instead, will compute the corrections to the constants given in the center of figure frame and shift them to the center of mass and principal axes.  

To account for the mass distribution we specify a vector offset from the center of mass to the center of figure, denoted as $\bfm{R}$, and a transformation dyad from the figure principal axes to the true principal axes, $\bfm{T}$ (see \cite{scheeres_YORP} for a brief discussion of dyads as used in this theory).  Thus, the new moment, accounting for the mass offset, is computed as:
\beq
	\bfm{M}_{CM} & = & \bfm{M}_{CF} + \bfm{R}\times\bfm{F}_{CF}
\eeq
where we introduce the $CM$ notation for center of mass and $CF$ for center of figure.  Note that $\bfm{F}_{CF} = \bfm{F}_{CM}$, and thus that $\bfm{A}_{l,CF} = \bfm{A}_{l,CM}$ and $\bfm{B}_{l,CF} = \bfm{B}_{l,CM}$, and so we suppress the $CM$ and $CF$ notation for the force coefficients.  Carrying these changes over to the moment Fourier coefficients we find:
\beq
	\bfm{C}_{l,CM} & = & \bfm{C}_{l,CF} + \bfm{R}\times\bfm{A}_{l} \\
	\bfm{D}_{l,CM} & = & \bfm{D}_{l,CF} + \bfm{R}\times\bfm{B}_{l} 
\eeq
To map these vectors into the principal axis frame we just pre-multiply with $\bfm{T}$.

The corrected equations for the constants in Eqns.\ \ref{eq:ommsecE} - \ref{eq:opmsecE} are then computed as:
\beq
	C_{0,z,CM} & = & \hat{\bfm{z}} \cdot \bfm{T} \cdot \left[ \bfm{C}_{0,CF} + \bfm{R}\times\bfm{A}_0 \right]  \label{eq:C0zCM} \\
	C_{1,x,CM} & = & \hat{\bfm{x}} \cdot \bfm{T}\cdot\left[ \bfm{C}_{1,CF} +  \bfm{R}\times\bfm{A}_1  \right] \label{eq:C1xCM} \\
	C_{1,y,CM} & = & \hat{\bfm{y}} \cdot \bfm{T}\cdot\left[ \bfm{C}_{1,CF} + \bfm{R}\times\bfm{A}_1  \right] \label{eq:C1yCM} \\
	D_{1,x,CM} & = & \hat{\bfm{x}} \cdot \bfm{T}\cdot\left[ \bfm{D}_{1,CF} + \bfm{R}\times\bfm{B}_1  \right] \label{eq:D1xCM} \\ 
	D_{1,y,CM} & = & \hat{\bfm{y}} \cdot \bfm{T}\cdot\left[ \bfm{D}_{1,CF} + \bfm{R}\times\bfm{B}_1   \right] \label{eq:D1yCM} 
\eeq
We should recall that all of the above coefficients are actually functions of the obliquity.  In keeping with previous practice, we compute these coefficients at discrete values of obliquity, although it is possible to define a unique series for each coefficient in the obliquity, as explicitly shown in \cite{CMDA_YORP}.

If we assume that the degree of density inhomogeneity is relatively small we can develop explicit corrections.  Then the magnitude of the center of figure offset from the center of mass is small relative to the mean radius of the body.  
The implications of this assumption for the re-orientation of the body axes can be seen by decomposing the reorientation dyad using axis-angle variables:
\beq
	\bfm{T} & = & \cos\phi\bfm{U} + (1 - \cos\phi) \hat{\bfm{a}} \hat{\bfm{a}} + \sin\phi\tilde{\bfm{a}} 
\eeq
where $\bfm{U}$ is the identity dyad, $\hat{\bfm{a}}$ is the unit vector pointing along the axis of rotation for the re-orientation, $\phi$ is the rotation angle about that axis, and the notation $\tilde{\bfm{a}}$ denotes the cross-product dyad applied to the vector $\bfm{a}$.  Our smallness assumption for the re-orientation means that both the angle $\phi$ is small, and that the axis of rotation is close to the original $z$-axis, or that to first order the axis equals $\hat{\bfm{a}} = a_x \hat{\bfm{x}} + a_y\hat{\bfm{y}} + \hat{\bfm{z}}$.  This vector has unit magnitude as $a_x^2$ and $a_y^2$ can be ignored as being small.  Ignoring higher-order combinations of $\phi$, $a_x$ and $a_y$ yields the first order result
\beq
	\bfm{T} & \sim & \bfm{U} + \phi \tilde{\hat{\bfm{z}}} + \ldots
\eeq
with the main effect being the shift of the $x$-$y$ axes about the nominal rotation pole.

Applying these smallness assumptions to Eqns.\ \ref{eq:C0zCM}-\ref{eq:D1yCM} yields the explicit corrections: 
\beq
	C_{0,z,CM} & = & C_{0,z,CF} + R_x A_{0,y} - R_y A_{0,x} \\
	C_{1,x,CM} & = & C_{1,x,CF} + R_y A_{1,z} - R_z A_{1,y} \nonumber \\
	& & + \phi\left[  C_{1,y,CF} + R_z A_{1,x} - R_x A_{1,z} \right] \\
	C_{1,y,CM} & = & C_{1,y,CF} + R_z A_{1,x} - R_x A_{1,z} \nonumber \\
	& & - \phi\left[ C_{1,x,CF} + R_y A_{1,z} - R_z A_{1,y}  \right] \\
	D_{1,x,CM} & = & D_{1,x,CF} + R_y B_{1,z} - R_z B_{1,y} \nonumber \\
	& & + \phi\left[  D_{1,y,CF} + R_z B_{1,x} - R_x B_{1,z} \right] \\
	D_{1,y,CM} & = & D_{1,y,CF} + R_z B_{1,x} - R_x B_{1,z} \nonumber \\
	& & - \phi\left[ D_{1,x,CF} + R_y B_{1,z} - R_z B_{1,y}  \right] 
\eeq
In the above formulae we have ignored second order combinations of $\phi$ and $R_x$ and $R_y$.  

Finally, let us consider the moment of inertia about the largest principal axis, $I_{z,CM}$.  Following from standard results this is computed as:
\beq
	I_{z,CM} & = & \hat{\bfm{z}} \cdot \bfm{T} \cdot \left[ \bfm{I}_{CF} + M \tilde{\bfm{R}}\cdot\tilde{\bfm{R}} \right] \cdot \bfm{T}^T \cdot \hat{\bfm{z}}
\eeq
From the first order dyad correction term we note that $\hat{\bfm{z}}\cdot\bfm{T} = \hat{\bfm{z}}$, as the first order correction terms are orthogonal to $\hat{\bfm{z}}$.  Thus we arrive at the updated moment of inertia, correct to the second order in the re-orientation terms:
\beq
	I_{z,CM} & = & I_{z,CF} - M \left( R_x^2 + R_y^2\right)
\eeq
If we consider the mass offset terms to be small the moment of inertia is essentially unchanged.

\section{Representative Corrections for Itokawa}

By computing the correction terms related to the force coefficients we can gauge the potential strength of this effect independent of a measurement of the mass offset.  We perform the computation for the asteroid Itokawa.  We will not focus on the obliquity corrections, as they will not be directly measurable, but focus on the spin-rate correction term coefficients $A_{0,x}$ and $A_{0,y}$.

Shown in Fig.\ \ref{fig:YORP_coef} is the YORP spin rate torque coefficient $C_{0,z,CF}$ and in Fig.\ \ref{fig:YORP_Fcoef} the averaged force coefficients, all as a function of obliquity (the current value of Itokawa obliquity is approximately $178^{\circ}$).  
In all of these plots we show results from a range of surface resolution models.  This is to emphasize the fact that the YORP effect is sensitive to the asteroid shape, and that it evidently has not converged even at the 0.5 meter resolution shape model \cite{scheeres_LPSC}.
We only compute the emission terms, as these are expected to be the dominant contributor.  

For Itokawa's obliquity of $178^\circ$ we note that $C_{0,z,CF}$, $A_{0,x}$ and $A_{0,y}$ are all negative.  Thus Itokawa should currently be experiencing a deceleration in its spin rate.  A positive shift in the center of mass from the center of figure in the $x$ axis (or $R_x < 0$) will cause an increase in the acceleration, making the body decelerate less rapidly for a small enough shift, and a positive $y$ axis shift in the center of mass ($R_y < 0$) will cause a decrease in the YORP rate of the asteroid, making it decelerate faster.  A convenient way to measure the strength of this effect is to compute the necessary offset to cause the YORP spin rate of change to equal zero.  Using the 4 meter resolution model we note that a shift of 11 meters in the center of mass from the center of figure along the $+x$-axis or a shift of 15 meters along the $-y$-axis would cause the YORP rate to be zero. For the 0.5 meter resolution model we find that the zero line corresponds to a $+x$-axis offset of 19 meters and a $-y$-axis offset of 30 meters.  This can be graphically seen by superimposing the zero-YORP rate location of the center of mass onto the pole projection of Itokawa, centered at the center of figure, shown in Fig.\ \ref{fig:CM_line}.  The smallest shift in the center of mass that can cause a zero YORP effect is 16 meters for the 0.5 meter resolution shape model and less than 9 meters for the 4 meter model.  

In \cite{Science_fujiwara, Science_yano} and \cite{Science_miyamoto} detailed descriptions of the morphology of the Itokawa ``neck'' region are given which find that loose regolith has migrated into this region from neighboring locations with higher elevation.  The remaining surface of the asteroid is found to be rather rocky and evidently devoid of a finer regolith covering.  Assuming a common grain density for all of the material, we would expect that the neck region would then have a higher bulk density than the surrounding regions.  This would cause a shift in the center of mass from the center of figure towards the neck region, which is consistent with an increase in the YORP deceleration rate, giving it a smaller magnitude, and potentially even giving it a positive rotation rate.  This also provides an explanation for why the YORP deceleration of Itokawa has not been detected as of yet.

\section{Discussion and Conclusions}

The inclusion of density inhomogeneities in the Itokawa mass distribution, with a concentration of density in the neck region of that body, may account for its non-detection, to date, and implies that its detection epoch will be delayed.  An implication of this result is the possibility to constrain the mass distribution of Itokawa once its YORP rotational acceleration is measured.  Then the center of mass of the body can be constrained to lie on lines similar to those shown in Fig.\ \ref{fig:CM_line}, as a function of which shape model is used.  Given such results it would then be possible to investigate different assumed regolith depths in the neck region to constrain the observed shift.

This result is currently only applicable to the asteroid Itokawa, as it is the only asteroid which should be affected by YORP and for which we have a precise shape, spin pole and mass estimate.  Future missions are currently being planned and proposed to asteroids, some of which should be affected by YORP.  This theory should be of especial interest, as measurements of the shape, mass and YORP effect can then be used to shed additional insight into an asteroid's mass distribution.

\clearpage

\section*{Figure Captions}

\begin{figure}[h!]
\caption{YORP torque coefficients $C_{0,z,CF}$ for Itokawa at three different surface resolutions.}
\label{fig:YORP_coef}
\end{figure}

\begin{figure}[h!]
\caption{YORP force coefficients a) $A_{0,x}$ and b) $A_{0,y}$ for Itokawa at three different surface resolutions.}
\label{fig:YORP_Fcoef}
\end{figure}

\begin{figure}[h!]
\caption{Zero deceleration center of mass line projected onto a pole-on view of Itokawa.  If the center of mass is to the right of this line, Itokawa will be accelerating due to YORP.  If it is to the left of this line, it will be decelerating.}
\label{fig:CM_line}
\end{figure}

\clearpage

\begin{figure}[h!]
\centering

\includegraphics[scale=0.75]{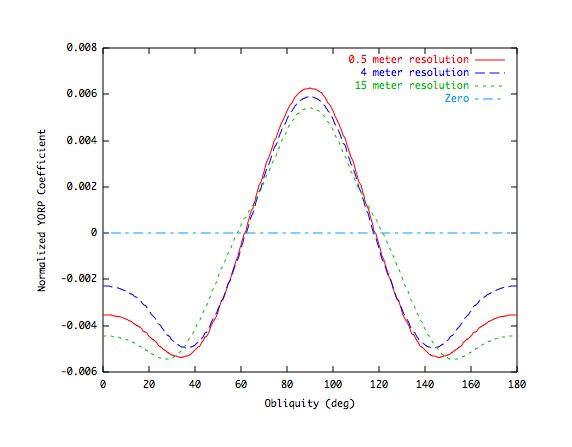}
%\caption{YORP torque coefficients $C_{0,z,CF}$ for Itokawa at three different surface resolutions.}
%\label{fig:YORP_coef}

\end{figure}
Figure \ref{fig:YORP_coef}

\clearpage

\begin{figure}[h!]
\centering

\includegraphics[scale=0.75]{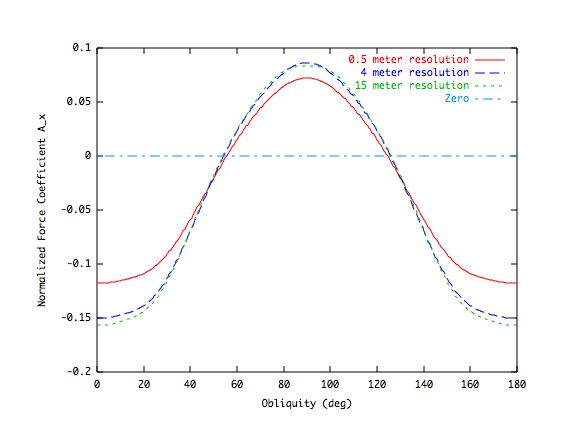}
\end{figure}
Figure \ref{fig:YORP_Fcoef} a

\clearpage

\begin{figure}[h!]
\centering

\includegraphics[scale=0.75]{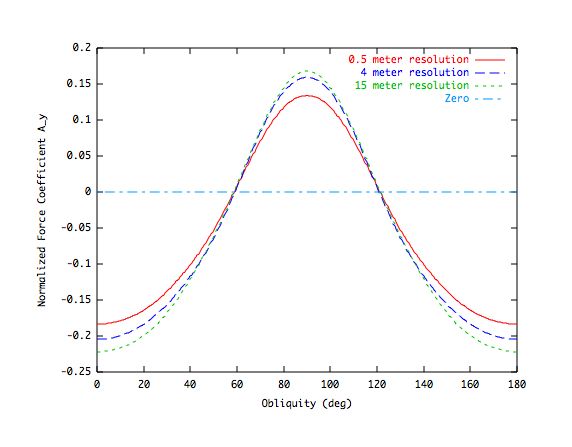}
%\caption{YORP force coefficients a) $A_{0,x}$ and b) $A_{0,y}$ for Itokawa at three different surface resolutions.}
%\label{fig:YORP_Fcoef}
\end{figure}
Figure \ref{fig:YORP_Fcoef} b

\clearpage

\begin{figure}[h!]
\centering

\includegraphics[scale=1.]{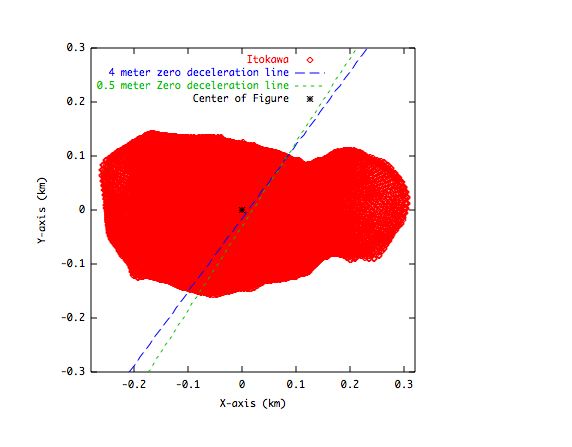}
%\caption{Zero deceleration center of mass line projected onto a pole-on view of Itokawa.  If the center of mass is to the right of this line, Itokawa will be accelerating due to YORP.  If it is to the left of this line, it will be decelerating.}
%\label{fig:CM_line}

\end{figure}
Figure \ref{fig:CM_line} 

\end{document}